# A Review on Cooperative Adaptive Cruise Control (CACC) Systems: Architectures, Controls, and Applications

Ziran Wang, *Student Member, IEEE*, Guoyuan Wu, *Senior Member, IEEE*, and Matthew J. Barth, *Fellow, IEEE*

*Abstract*—Connected and automated vehicles (CAVs) have the potential to address the safety, mobility and sustainability issues of our current transportation systems. Cooperative adaptive cruise control (CACC), for example, is one promising technology to allow CAVs to be driven in a cooperative manner and introduces system-wide benefits. In this paper, we review the progress achieved by researchers worldwide regarding different aspects of CACC systems. Literature of CACC system architectures are reviewed, which explain how this system works from a higher level. Different control methodologies and their related issues are reviewed to introduce CACC systems from a lower level. Applications of CACC technology are demonstrated with detailed literature, which draw an overall landscape of CACC, point out current opportunities and challenges, and anticipate its development in the near future.

## I. Introduction

Rapid development of our transportation systems has brought us much convenience in daily life, allowing both passengers and goods to be transported domestically and internationally in a quicker fashion. It is estimated that more than one billion motor vehicles are owned by people over the globe, and this number will be even doubled within one or two decades [1]. A series of issues related to our transportation systems are generated by such a huge number of motor vehicles. In term of safety, there are more than 30,000 people perish from roadway accidents on U.S. highway every year [2]. In terms of mobility, Los Angeles in U.S., for instance, tops the global ranking with 104 hours spent per commuter in traffic congestions during 2016 [3]. And in terms of sustainability, 3.1 billion gallons of energy were wasted worldwide due 7to traffic congestion [4].

Significant developments of connected and automated (CAV) technology have been achieved during the last decade, addressing aforementioned issues of our current transportation systems. Both connectivity and automation are integrated in CAVs, making them capable to not only drive by themselves with on-board sensing sensors, but also communicate with each other by vehicle-to-vehicle (V2V) communications. Cooperative adaptive cruise control (CACC) is one of the most promising technologies for CAVs, which extends adaptive cruise control (ACC) with cooperative maneuvers by CAVs.

In CACC systems, CAVs share their own parameters with other CAVs in the network by V2V communications, which is realized in autonomous manner without central management [5]. Given the fact that the communication bandwidth might become insufficient when the number of CAVs increases in a CACC system, short ranged wireless technologies are more accepted for V2V communications [6]. So far, the most dominant V2V communication protocol is Dedicated Short Range Communications (DSRC), and other advanced communication protocols have also been proposed and developed for V2V communications, such as LTE and 5G [7], [8].

CACC takes advantage of V2V communications to allow CAVs to form platoons and be driven at harmonized speeds with shorter time headways between them. By sharing vehicle information such as acceleration, speed, and position in a distributed manner, CAVs in a certain communication range can cooperate with others to obtain the following benefits: 1) driving safety is increased since actuation time is shortened compared to manually driven, and downstream traffic can be broadcasted to following vehicles in advance; 2) roadway capacity is increased due to the reduction of time/distance headways between vehicles; 3) energy consumption and pollutant emissions are reduced due to the reduction of unnecessary velocity changes and aerodynamic drag on following vehicles.

The remainder of this paper is organized as follows: Chapter II reviews literature related to the architecture of CACC systems. Chapter III focuses on the control aspect of CACC systems, where different literature is reviewed by their category. Applications of CACC technology are introduced with detailed examples in Chapter IV. Chapter V concludes the paper with some further discussions.

## II. Architectures

It was stated in UC Berkeley PATH's former research that current CACC implementations in production vehicles are mainly developed as the extension of commercially available adaptive cruise control (ACC) systems [9]. Therefore, most CACC vehicles are also equipped with sensors installed on

Ziran Wang is with Department of Mechanical Engineering, Guoyuan Wu and Matthew J. Barth are with Department of Electrical and Computer Engineering, and they are all with the College of Engineering – Center for Environmental Research and Technology (CE-CERT), University of California, Riverside, CA 92507, USA. Email: zwang050@ucr.edu, gywu@cert.ucr.edu, barth@ece.ucr.edu.

ACC vehicles, such as odometer, radar and/or Lidar. In their work, the overall system structure of a CACC-enabled CAV is illustrated as Fig. 1. A similar schematic representation was also proposed by Ploeg *et al.* in their design and experimental evaluation of CACC [10].

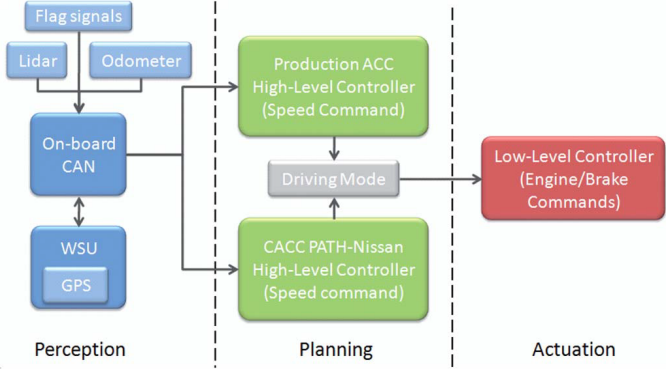

Fig. 1. System structure of a CACC-enabled CAV.

In the perception phase, a CAV gets information from the sensors installed on itself, and includes them on the CAN bus data structure. Specifically, information received from the wireless safety unit (WSU) consists of two parts: 1) data transmitted by other CAVs in the CACC system through V2V communications such as speed, acceleration, inter-vehicle distance, current time gap and so on; 2) data collected by GPS with wider area augmentation system differential corrections including detection and assignment of the vehicle position sequence in the CACC system. On the other hand, information derived from on-board sensors such as Lidar, odometer and flag signals will also be received and included on the CAN bus data structure.

The planning phase includes the high-level controller, and this is where researchers and engineers propose and implement vehicle longitudinal control algorithms. In this study, the commercial ACC controller is already installed on the CAV, while the CACC controller code is developed in MATLAB/Simulink and loaded in the CAV using a dSpace MicroAutoBox, which is connected to the vehicle via the CAN bus. The ACC system and CACC system are both available and the driver can switch between them in real time. The actuation phase is in charge of executing target reference command transmitted from the planning phase, where the low-level controller converts the target speed commands into throttle and brake actions.

Although the system structures of CAVs in different CACC systems are somewhat similar to Fig. 1, the vehicle information flow topology varies based on different methodologies. The information flow topology defines the way a CAV obtains information from other CAVs in a CACC system. Some typical types of information flow topologies include predecessor-following, predecessor-leader following, two predecessor-following, two predecessor-leader following and bidirectional types, which are illustrated in Fig. 2 [11].

Early-stage CACC systems are mainly based upon on-board sensors such as radar to detect surrounding environment, which means a vehicle in the system can only obtain information from its immediate predecessor or follower. Therefore, predecessor-following and bidirectional information flow topology are typical choices for those CACC systems [12]. However, taking advantage of V2V communications, recently proposed CACC systems allow CAVs to transmit information among each other in a much wider range, and that's how information flow topologies like predecessor-leader following, two predecessor following and two predecessor-leader following are developed.

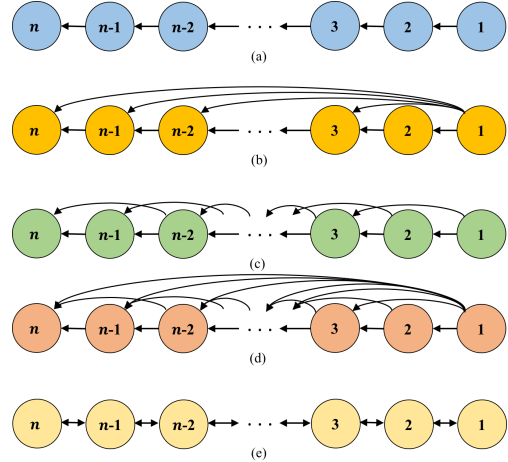

Fig. 2. Typical information flow topologies: (a) predecessor-following, (b) predecessor-leader following, (c) two predecessor-following, (d) two predecessor-leader following, and (e) bidirectional.

### III. CONTROLS

The vehicle control strategy plays a significant role in the CACC system, since the vehicle dynamics is calculated by the vehicle controller. More specifically, the longitudinal control strategy has been heavily studied by a variety of research, since vehicles in a CACC system needs to maintain the same longitudinal speed with the other vehicles in the system, while keeping a constant longitudinal inter-vehicle distance/headway with respect to its preceding vehicle. Different longitudinal controllers have been proposed to address different issues, such as forming platoons, optimizing fuel consumption, or ensuring system stability. In the following subsections, different longitudinal controllers including the most recently proposed ones are analyzed and categorized into separate strategies.

#### A. Model Predictive Control

Model predictive control (MPC) refers to a class of control algorithms that utilize an explicit process model to predict the future response of a plant [13]. Traditionally, MPC is formulated in the state space for a single-agent system, where the system to be controlled is described by a linear discrete time model as

$$x(k+1) = Ax(k) + Bu(k), x(0) = x_0 \qquad (1)$$

where $x(k) \in \mathbb{R}^n$ denotes the state input, and $u(k) \in \mathbb{R}^m$ denotes the control input [14]. Typically, a receding horizon implementation will be formulated by the open-loop

optimization problem as

$$J_{(p,m)}(x_0) = \min_{u(\cdot)}[x^T(p)P_0 x(p) + \sum_{i=0}^{p-1} x^T(i)Qx(i) + \sum_{i=1}^{m-1} u^T(i)Ru(i)] \quad (2)$$

subject to

$$Ex + Fu \leq \varphi \quad (3)$$

where $p$ denotes the length of the prediction horizon, and $m$ denotes the control horizon [15].

Stanger *et al.* applied the linear MPC approach to a CACC system, where the fuel consumption is minimized by including the nonlinear static fuel consumption map of the internal combustion engine into the control design with a piecewise quadratic approximation [16]. A similar stochastic linear MPC-based CACC control strategy was developed by Moser *et al.*, where a simulation study is conducted using IPG CarMaker and MATLAB and shows 11~15% fuel savings of the proposed strategy [17]. Lang *et al.* made use of MPC together with V2V and V2I communications to achieve an accurate prediction of preceding vehicle's velocity trajectory [18], [19]. It is concluded by their work that the achievable fuel benefits of the proposed MPC-based CACC system depend on the road segment, the allowed minimum and maximum inter-vehicle distance to the preceding vehicle, and the quality of the velocity prediction.

Generally, traditional MPC systems that are implemented in a centralized manner assume all states are known to compute control inputs. However, many centralized applications are not that practical in some scenarios of traffic system, due to the limitations to gather information from all agents and infrastructures, and to compute large-scale optimization problems. Therefore, distributed MPC (DMPC) schemes have been proposed to address that issue.

In DMPC, each controller adopts aforementioned MPC strategy for controlling its system, where it not only considers dynamics, constraints, objectives, and disturbances of the subsystem, but also considers the interactions among different systems [20]. Local controllers solve their MPC problems based on local information and also share information with others to improve the overall performance.

Recently, many researchers have been focused on proposing DMPC schemes for cooperative operational systems. The majority of existing DMPC algorithms aim to stabilize systems with a common set point, assuming all agents know the desired equilibrium information. However, some also study the asymptotic stability by employing the consistency constraints, where both the newly and previously calculated optimal trajectories should be bounded. Dunbar *et al.* considered the case when state vectors of interacting systems are coupled in a single cost function of a finite horizon optimal problem, but dynamics and constraints are decoupled [21]. Keviczky *et al.* considered the case when cost function and constraints couple the dynamical behavior of a set of dynamically decoupled systems [22].

### B. Consensus Control

Consensus problem has been developed and analyzed by researchers in the field of distributed computing for a long history. Essentially, consensus means a network of agents cooperatively reach an agreement with respect to a certain interest which depends on the states of all agents. Instead of being controlled by a centralized scheme, which assumes the global team knowledge is available to all agents in the network, consensus can act as a distributed scheme which requires only local interactions and evolves in a parallel manner [23].

Basically, the distributed consensus algorithm is designed to impose similar dynamics on the states of agents in the network. If the communication bandwidth is sufficiently large, the state updates of vehicles in the network can be modeled by differential equations. Otherwise, if the communication data are discrete, the state updates are modeled by difference equations [24]. The most common continuous consensus algorithm [25]–[27] is a single-integrator algorithm and can be given by

$$\dot{x}_i(t) = -\sum_{j=1}^{n} a_{ij}(t)(x_i(t) - x_j(t)), i = 1, \ldots, n \quad (4)$$

where $a_{ij}(t)$ is the $(i,j)$ entry of the adjacency matrix of the associated communication graph of the system at time $t$, $x_i$ is the information state of the $i$th agent. When the information state $x_i(t)$ of the $i$th agent is driven toward the information states of its neighbors, the consensus of Eq. (4) is reached.

The single-integrator distributed consensus algorithm Eq. (4) can be extended to a double-integrator algorithm [28], [29] to model the dynamics of physical agents such as CAVs in a CACC system. The double-integrator distributed consensus algorithm can be given by

$$\ddot{x}_i(t) = -\sum_{j=1}^{n} a_{ij}(t)[(x_i(t) - x_j(t)) + \gamma(\dot{x}_i(t) - \dot{x}_j(t))], i = 1, \ldots, n \quad (5)$$

where $\gamma > 0$ is a tuning parameter that denotes the coupling strength between the state derivatives. The double-integrator distributed consensus system can be applied to tackle formation control problems of vehicles [30], where $x_i(t)$, $\dot{x}_i(t)$, and $\ddot{x}_i(t)$ can be considered as the (longitudinal and/or latitudinal) position, velocity and acceleration of vehicle $i$ at time $t$, respectively. The acceleration of vehicle $i$ is calculated on the basis of the velocity and acceleration of itself and its neighboring vehicles. Consensus of this system is reached when both position consensus $(x_i(t) - x_j(t))$ and velocity consensus $(\dot{x}_i(t) - \dot{x}_j(t))$ are reached by vehicles in the network.

The issue of time delay appears in most urban traffic systems due to several reasons, including sensing delay, communication delay, computation delay, and actuation delay. Since time delay possibly degrades the performance and stability of a system, studies about distributed consensus algorithm have been conducted to include time delay in the system [31]. The single-integrator distributed consensus algorithm with time delay [32], [33] can then be categorized

as

$$\dot{x}_i(t) = -\sum_{j=1}^{n} a_{ij}(t)(x_i(t-\tau) - x_j(t-\tau)), i = 1, \dots, n \quad (6)$$

and the double-integrator distributed consensus algorithm with time delay [34]–[37] can be categorized as

$$\ddot{x}_i(t) = -\sum_{j=1}^{n} a_{ij}(t)[(x_i(t-\tau) - x_j(t-\tau)) + \gamma(\dot{x}_i(t-\tau) - \dot{x}_j(t-\tau))], i = 1, \dots, n \quad (7)$$

where $\tau$ denotes the time-variant or time-invariant delay.

Although different consensus control algorithms have been proposed and evaluated for CACC systems, there are several disadvantages of such a linear design methodology. For instance, linear methodologies like consensus control cannot deal with nonlinearity and constraints of the CACC system. Therefore, other approaches have been proposed to handle these issues, such as the optimal control approach which will be discussed in the next subsection. Additionally, it is not that easy to address the string stability issue explicitly, where extra requirements must be fulfilled such as the information flow topology, or the spacing policy. The details of string stability will be covered in the last subsection of this chapter.

### C. Optimal Control

Besides MPC mentioned in the earlier subsection, other optimal control methodologies are discussed in this subsection. Optimal control has been considered as an approach for CACC systems by many research work. In general, the design of optimal controller for CACC systems can be equivalently formulated as a structured convex optimization problem with the objective to minimize energy consumption or travel time. Unlike most consensus control approaches that only consider vehicle speed and position as inputs, and simplify the longitudinal control of a CACC system to a single integrator or double integrator problem based on the assumption of linearity, optimal control approaches often take nonlinearity and constraints into account, such as vehicle powertrain and vehicle aerodynamics.

In many cases that optimal control approach is applied to CACC systems, total energy consumed by vehicles traversing the designated area acts as the objective function. In the Eco-CACC system proposed by Yang et al. [38], the optimal control problem is defined as

$$\min_{a_-, a_+} \int_{t_0}^{t_0+T} F(v(t), v'(t)) dt, \quad (8)$$

$$s.t., \int_{t_0}^{t_0+T} v(t) dt = d + l, \quad (9)$$

$$0 \le a_- \le a_-^s, \quad (10)$$

$$0 \le a_+ \le a_+^s. \quad (11)$$

$F(\cdot, \cdot)$ is defined as a nonlinear function of speed $v(t)$ and acceleration $v'(t)$ to estimate the energy consumption rate based on vehicular speed and acceleration levels. It also subjects to some constraints of speed and maximum acceleration. Zohdy et al. proposed an optimal control-based intersection CACC (iCACC) system, and the simulation results show the proposed system reduces the average intersection delay by 90% and energy consumption by 45%, respectively [39]. Van de Hoef et al. formulated a combinatorial optimization problem to maximize the fuel savings of coordination leaders of a Truck CACC system, and furtherly formulated a convex optimization problem with linear constraints for a group consisting of a coordination leader and its coordination followers [40].

Wang et al. proposed a platoon-wide Eco-CACC system, aiming to minimize the platoon-wide energy consumption and pollutant emissions at different stages of the CACC operation [41]. A further study about the intra-platoon vehicle sequence was conducted by optimization methodology [42]. Jovanovic et al. developed centralized linear quadratic optimal control formulations to penalize relative position errors between neighboring CACC vehicles, concluding stability and detectability will decrease when the size of the CACC system increases [43]. Turri et al. studied the cooperative look-ahead control of a heavy-duty CACC system, where the fuel model is formulated as an optimal control problem to find the optimal engine speed to minimize fuel consumption [44].

### D. String Stability

String stability is a basic requirement to ensure the safety of a CACC system. Specifically, string stability is a desirable characteristic for platoons to attenuate either distance error, velocity or acceleration along upstream direction. The problem can be formulated as

$$\|y\|_\infty \le \|u\|_\infty \quad (12)$$

where $y$ is the scalar output of either distance error, velocity or acceleration of the following vehicle $i+1$, and $u$ is the scalar output of either distance error, velocity or acceleration of the preceding vehicle $i$. Then, string stability is guaranteed if

$$\left\|\frac{Y(s)}{U(s)}\right\|_\infty \le 1 \quad (13)$$

where $Y(s)$ and $U(s)$ are the Laplace transforms of $y$ and $u$.

The internal Lyapunov stability of a platoon in CACC systems does not necessarily lead to string stability. If the string stability cannot be guaranteed, error signals will be amplified along upstream direction even if the closed-loop system is internal stable, eventually resulting in collision of two consecutive vehicles. String stability should be considered when any novel CACC system is proposed. A reinforcement learning-based CACC system was proposed in [45], which modeled the system as a Markov Decision Process incorporated with stochastic game theory. The string stability was proved where the proposed system was capable of damping small disturbances throughout the platoon. Padé approximation method was adopted in the CACC system proposed by Xing et al. to model the vehicle actuator delay arriving at a finite-dimensional model, since the vehicle actuator delay significantly limits the minimum inter-vehicle distance in view of string stability requirements [46].

Many other works also focused on analyzing the string stability of CACC systems [62], [77]-[85], and the following conclusions regarding ensuring string stability have been proposed: 1) If the constant distance spacing policy for vehicles in the platoon is adopted, a predecessor-following information flow topology cannot guarantee string stability. Broadcasting leader's information to following vehicles in the platoon through V2V communications can extend the information flow, and thus ensure string stability. 2) Instead of adopting a constant distance spacing policy, constant time headway spacing policy can be used to ensure string stability, where the inter-vehicle distance relies on the velocity of vehicle, and therefore relax the formation rigidity of the system.

IV. APPLICATIONS

CACC technology introduces benefits to current transportation systems with respect to safety, mobility and sustainability. Applications of CACC technology have been proposed and developed over the years under different traffic networks, and field implementations of such applications have also been conducted to test the effectiveness of CACC technology.

*A. Vehicle Platooning*

CACC technology allows CAVs to form vehicle platoons with shorter inter-vehicle distances. Since vehicles are tightly coupled with their neighbors, the roadway capacity is highly increased, while the energy consumption is reduced due to the mitigation of aerodynamic drags and unnecessary speed changes. Numerous research work has been conducted so far to apply CACC technology to vehicle platooning.

As mentioned in the previous chapter, distributed consensus strategy has been widely adopted to design the longitudinal controller of the vehicle platooning application for its easiness of comprehensive theoretical analysis. Wang *et al.* developed first-order distributed consensus algorithms to achieve weighted and constrained consensus for inter-vehicle distance in the platoon [47], [48]. Second-order distributed consensus algorithms attracted more attention with respect to first-order, where Wang *et al.* [49], [50] adopt predecessor-following information flow topology for vehicles in the platoon, while Di Bernardo *et al.* [51], [52] and Jia *et al.* [53] used predecessor-leader following topology. Third-order consensus algorithms are discussed in [54], [55], where the longitudinal acceleration difference of two vehicles was also taken into account for vehicle longitudinal control.

Since the difficulty of distributed consensus strategy to deal with nonlinear systems, MPC acts as a substitute strategy to control CACC-enabled vehicles in vehicle platooning application with nonlinearities by formulating multiple local convex issues in a predictive horizon [56]. MPC was firstly adopted to develop ACC system in [57], [58], where only two vehicles in the formation are involved. Researchers extended the use of MPC from ACC to CACC in [59], [60], considering not only the predecessor-following information flow topology, but also predecessor-leader following, two-predecessor following, and others.

Many other cooperative distributed control strategies for vehicle platooning using CACC technology have also been proposed and evaluated, allowing vehicles to cooperate with each other in a distributed manner while reaching a global goal [9], [10], [61]-[66]. Specifically, the following issues were or should be addressed in those work while designing vehicle platooning applications:

*a)* Wireless communication issues among vehicles. Wireless V2V communications inevitably introduce network imperfections, such as communication delay, communication range limit, packet loss, and sampling intervals. Among proposed cooperative distributed control strategies, most of them have taken communication issues into account. Specifically, Di Bernardo *et al.* considered heterogeneous time-varying communication delays in their control algorithms as stochastic variables with a uniform discrete distribution [51], [52]. Jia *et al.* used the network simulator OMNeT/MiXiM to simulate the 802.11p standard-based V2V communications in the proposed system, where propagation delay, transmission range, beacon frequency, beacon size, and other communication issues were taken into account [53].

*b)* Impact of vehicle platooning on real traffic situations. Although most proposed control strategies have been mathematically proven in theory and analyzed in simulation software, relatively few of them have conduct experiment to test their strategies in real-world situations. Milanés *et al.* implemented four production Infiniti M56s vehicles with 5.9-GHz DSRC to validate the performance of the proposed controller in gap setting changes test, cut-in and cut-out test, and four-vehicle test [9]. Ploeg *et al.* selected Toyota Prius III Executive as their model to demonstrate the technical feasibility of the proposed CACC control methodology [10]. The Grand Cooperative Driving Challenge (GCDC) held in Helmond in Netherlands in 2011 was aiming to support and accelerate the introduction of cooperative and automated vehicles in everyday traffic, which inspired many excellent research experiments in CACC systems [59], [85], [87].

*B. Eco-Driving on Signalized Corridors*

The cooperation between vehicles and intersections has been a popular topic in the research field of intelligent transportation systems for a long time. Since traffic signals are considered to be the most efficient way to control the traffic at intersections, numerous work have been conducted to increase their efficiency by cooperative operations. Specifically, how to integrate traffic signal information into CACC systems and hence reduce the overall energy consumption becomes a popular research topic. Many work referred this topic as "Eco-driving on signalized corridor", or "Eco-CACC".

Yang *et al.* developed an Eco-CACC algorithm for isolated signalized intersections, which computes the optimal vehicle trajectory to minimize the energy consumption by ensuring each vehicle arrive at the intersection as soon as the last vehicle in the queue is discharged [38]. In this work, vehicle queue at the intersection was analyzed by proposing a queue estimation model, and an energy optimization-based Eco-CACC algorithm was developed accordingly. Microscopic traffic simulation showed the proposed Eco-CACC system produces vehicle energy savings up to 40% when the CAV market penetration rate is 100%.

Zhody et al. proposed a heuristic game theory-based intersection CACC (iCACC) system and evaluated it by a case study comparing the proposed one with the baseline scenario, which is a four-way stop control [39]. They further improved the system which enables intersection controller to communicate with CAVs and give advices to them on the optimum course of action to minimize the intersection delay and energy consumption [67]. Another heuristic game theory-based iCACC system was developed by the same research group, where Malakorn et al. integrated CACC and intelligent traffic signal control technology and analyzed the mobility, energy and environmental impacts of the proposed system [68]. A simulation study was conducted, showing the expected benefits of implementing the proposed system would introduce 75% reduction in energy use.

Different from the aforementioned systems that integrate vehicles travelling through intersections into vehicle platoons, Wang et al. proposed a novel cluster-wise cooperative system along signalized arterials [69], [70]. All CAVs approaching a particular intersection will be grouped into different clusters with deterministic sequences based on their estimated time to arrive at the intersection. Each vehicle cluster consists of several CACC platoons on each lane, and different CACC platoons are coupled by the coordination among different platoon leaders. Therefore, when the cluster leader conducts eco-driving maneuver with respect to the traffic signal information, all the cluster followers are able to follow the movement of the cluster leader by the proposed CACC algorithm.

*C. Cooperative Merging at Highway On-Ramps*

Ramp metering has been considered as a commonly used application to regulate the upstream traffic flows on highway on-ramps. However, since it also introduces a stop-and-go scenario, which leads to extra energy consumption and time waste, many researchers have developed advanced methodology to address this issue. Specifically, CACC technology has been widely adopted to allow CAVs to merge with each other in a cooperative manner.

The concept of virtual vehicle or "ghost" vehicle of a CACC system in the highway on-ramps cooperative merging scenario was originated from Uno et al. [71]. The proposed approach maps a virtual vehicle onto the highway main road before the actual merging happens, allowing vehicles to perform safer and smoother merging maneuver. Lu et al. applied the similar idea in their proposed systems, where merging control algorithms were developed for automated highway systems [72]-[74]. Wang et al. proposed a distributed consensus-based highway on-ramp merging system, where two CACC systems are formed on the main road and on-ramp, respectively [75]. CAVs will be assigned with sequence IDs based on their arrival time at the merging point, and cooperate with their neighboring vehicles (either realistic ones on the same lane or "ghost" ones on the other lane) through V2V communications.

Other than the virtual vehicle concept, Milanés et al. proposed a fuzzy logic-based controller to control the longitudinal movement of vehicles in the proposed system, where the merging vehicle coming from the on-ramp needs to cooperatively merge with CACC vehicles on the main road [76]. Dao et al. proposed a distributed control protocol to assign vehicles into CACC systems in the merging scenario [77]. Scarinci et al. reviewed some of the other CACC-related control methodologies for improving on-ramp merging [78]. Rios-Torres et al. also reviewed some of the CACC or platoon-based strategies used by CAVs to merge at highway on-ramps [79].

V. DISCUSSIONS

This paper presents a review on CACC systems from perspectives of high-level system architectures, low-level control methodologies, and overall system applications. Although many theoretical and/or experimental results have been already provided in the paper, there are still some open questions that need to be addressed in future work. Based on the topics of our paper, we can briefly name a few of the issues:

*a)* How to build a more reliable architecture for CACC systems? Unlike most proposed CACC systems that assume a rather fixed environment, the realistic traffic network will introduce highly dynamic environment, including changing information flow topologies, varying workload distribution between different CAVs, and packet loss of V2V communications.

*b)* How to develop more ready-to-market control methodology for CACC systems? It is true that many advanced control methodologies have been proposed and tested in simulations, and some of them were even implemented in realistic vehicles. However, like most of our automated vehicles nowadays, such CAVs with predefined controllers still need to be tested under all kinds of different conditions and environments, and also for a rather long mileage. Since CACC systems often involves several CAVs, it would be relatively difficult to conduct enough tests of proposed systems before making them available on market.

*c)* Will the high cost impede the implementation of different CACC applications? Although different CACC applications have been proposed to address issues in our current transportation systems, showing some good results in terms of safety, mobility and sustainability, they are also based on many assumptions. Taking the cooperative merging on highway on-ramps application for example, it will definitely cost a lot of money for government to make new policies, update roadside infrastructure, test the proposed method in real traffic, and so on. To achieve a preferred penetration rate of CAVs in the application, the general public also need to spend money to purchase new vehicles. Such high costs may prevent CACC applications like this to be implemented in the future.